\documentclass[aps,shownopacs,superscriptaddress,nofootinbib,preprint,11pt]{revtex4}
\usepackage[section,subsection,subsubsection]{extraplaceins}
\usepackage{hyperref}
\usepackage{amsmath}
\usepackage{float}
\usepackage{graphics}
\usepackage{graphicx}
\usepackage{epsfig}
\usepackage{psfrag}
\usepackage{color}
\usepackage{slashed}

\usepackage{amsfonts}
\usepackage{amssymb}

\newcommand*{\be}{\begin{equation}}
\newcommand*{\ee}{\end{equation}}
\newcommand*{\bea}{\begin{eqnarray}}
\newcommand*{\eea}{\end{eqnarray}}

\newcommand{\comment}[1]{}

\newcommand{\cref}[1]{Chapter~\ref{c.#1}}

\def\beq{\begin{equation}}
\def\eeq{\end{equation}}
\def\bea{\begin{eqnarray}}
\def\eea{\end{eqnarray}}
\def\ba{\begin{array}}
\def\ea{\end{array}}
\def\bi{\begin{itemize}}
\def\ei{\end{itemize}}
\def\be{\begin{enumerate}}
\def\ee{\end{enumerate}}
\def\bc{\begin{center}}
\def\ec{\end{center}}
\def\bt{\begin{table}}
\def\et{\end{table}}
\def\btb{\begin{tabular}}
\def\etb{\end{tabular}}

\def\lsim{\raise0.3ex\hbox{$\;<$\kern-0.75em\raise-1.1ex\hbox{$\sim\;$}}}
\def\gsim{\raise0.3ex\hbox{$\;>$\kern-0.75em\raise-1.1ex\hbox{$\sim\;$}}}

\def\beq {\begin{equation}}
\def\eeq {\end{equation}}
\def\bea {\begin{eqnarray}}
\def\eea {\end{eqnarray}}

\preprint{CERN-TH-2016-002}
\preprint{TIFR/TH/16-01}
\begin{document}

\vspace*{1.cm}
\title{Kaluza-Klein gluon + jets associated production
	at the Large Hadron Collider}

\vspace*{0.5cm}
\author{ A. M. Iyer}
\email{abhishek@theory.tifr.res.in}
\affiliation{Department of Theoretical Physics, Tata Institute of Fundamental Research, Homi Bhabha Road, Colaba, Mumbai 400 005, India}
\author{ F. Mahmoudi}
\altaffiliation[]{Also Institut Universitaire de France, 103 boulevard Saint-Michel, 75005 Paris, France}
\email{nazila@cern.ch}
\affiliation{Univ Lyon, Univ Lyon 1, ENS de Lyon, CNRS, Centre de Recherche Astrophysique de Lyon UMR5574, F-69230 Saint-Genis-Laval, France}
\affiliation{Theoretical Physics Department, CERN, CH-1211 Geneva 23, Switzerland}
\author{N. Manglani}
\email{namratam@physics.mu.ac.in}
\affiliation{Department of Physics, University of Mumbai,
	Kalina, Mumbai 400098, India}
\affiliation{	Shah and Anchor Kutchhi Engineering College, Mumbai 400088, India.}
\author{K. Sridhar}
\email{sridhar@theory.tifr.res.in}
\affiliation{Department of Theoretical Physics, Tata Institute of Fundamental Research, Homi Bhabha Road, Colaba, Mumbai 400 005, India\vspace*{0.5cm}}

\begin{abstract}
The Kaluza-Klein excitations of gluons offer the exciting possibility of 
probing bulk Randall-Sundrum (RS) models.
In these bulk models either a custodial symmetry or a deformation of the metric away from AdS is invoked in order to deal 
with electroweak precision tests.
 Addressing both these models, we suggest a
new channel in which to study the production of KK-gluons 
($g_{KK}$): one where it is produced in association with one or more  hard
jets. The cross-section for the $g_{KK}+$ jets channel is significant because of several
contributing sub-processes. In particular, the 1-jet and the 2-jet associated processes
are important because at these orders in QCD the $qg$ and the $gg$ initial states respectively 
come into play.
We have performed a hadron-level simulation of the signal and present 
strategies to effectively extract the signal from what could potentially
be a huge background. We 
present results for the kinematic reach of the LHC Run-II for different 
$g_{KK}$ masses in bulk-RS models.

\end{abstract}

\maketitle
\section{The Bulk Randall-Sundrum Model}
\noindent The Randall-Sundrum model (RS model) \cite{Randall:1999ee} 
is a five-dimensional model with a warped metric and 
was proposed as a solution to the gauge-hierarchy problem. 
In this five-dimensional model, the fifth dimension $y$ is compactified
on an $S^1/Z^2$ orbifold of radius, $R$. At the fixed points of the orbifold,
$y=0$ and $y=\pi R \equiv L$, two branes--
the UV and the IR brane respectively are located. 

One starts with a warped metric given as:
\begin{equation}
	ds^2=e^{-2 A(y)}\eta_{\mu\nu}dx^{\mu}dx^{\nu}- dy^2
	\label{e1}
\end{equation}
and using a five-dimensional gravity action with a bulk cosmological
constant $\Lambda$ one can show that the solutions to the Einstein
equation imply 
\begin{equation}
	A(y) = \pm k \vert y \vert
	\label{e6.11}
\end{equation}
where $k^2 \equiv - \Lambda /12 M^3$ with $M$ being the Planck
scale. 

The Standard Model (SM) fields are all IR-localised and only
the graviton fields access the bulk.
The factor of $v / M \sim 10^{-16}$ (where $v$ is the vacuum expectation
value of the SM Higgs field) is generated by choosing a
value of $kL \sim 30$ thereby stabilising the gauge hierarchy. 

The problem, however, is that the suppression that one obtains for the 
Higgs vev materialises for all fields localised on the IR brane. 
Thus, mass scales which suppress dangerous
higher-dimensional operators responsible for proton decay or neutrino
masses also become small which spells a disaster for the RS model. One
way out of this is to realise that to solve the gauge-hierarchy problem
one needs to only localise the Higgs on the IR brane but all the other
SM fields could be in the bulk \cite{Davoudiasl:1999tf,Pomarol:1999ad, Gherghetta:2000qt, 
	Grossman:1999ra}. In fact, even the Higgs need not be
sharply localised on the IR brane but only somewhere close to it. In this
way, it is possible to make viable variations of the RS model, now 
collectively known as Bulk RS models. These models yield a bonus:
localising fermions at
different positions in the bulk gives different overlaps of their
profiles with the Higgs field, which is localised on or close to the 
IR brane. This gives rise in a natural way to the Yukawa-coupling hierarchy
\cite{Gherghetta:2000qt}.

The AdS/CFT correspondence when worked out for a slice of AdS spacetime
also provides both a motivation for and an understanding of bulk models.
(For a review, see Ref. \cite{Gherghetta:2010cj}.) In fact, this 
correspondence
shows that the dual of the RS model with all the SM fields on the IR brane
is a fully composite SM in four dimensions -- a theory which we know does not
survive in the face of extant experimental constraints. But when the SM fields
are localised at different positions in the bulk, one ends up with a 
partially composite SM where the compositeness is primarily in the Higgs, the
top and in the KK sector. Such a model is, in fact, a viable model as
we will discuss in the following.

It is possible to construct a bulk-extension of the SM by having
the gauge and fermion fields in the bulk, the Higgs localised on
or near the IR brane and with a suitable mechanism to make the
model successfully confront constraints from electroweak precision
measurements \cite{Gherghetta:2010cj}. Such a model has 
interesting features. Other than providing a framework for addressing
the question of fermion mass hierarchy, it also naturally
results in small mixing angles in the Cabibbo-Kobayashi-Maskawa (CKM)
matrix, gauge-coupling universality and suppression of 
flavour-changing neutral currents to experimentally acceptable values
\cite{Burdman:2003nt, Huber:2003tu, Casagrande:2008hr, Bauer:2009cf,
	Agashe:2004cp}.

Electroweak precision tests provide very strong constraints on bulk
models. If, for example, only gauge bosons propagate in the bulk but
the fermions are localised on the IR brane then the couplings of
the gauge boson Kaluza-Klein (KK) modes to the IR-localised fermions
yield unacceptably large contributions to $T$ and $S$ and this gives
a lower bound of 25 TeV on the mass of the first KK mode of the gauge boson. 
Of course, one way to relax this bound is to localise the fermions
in the bulk and especially the light fermions close to the UV brane
and this significantly reduces the constraint coming from the $S$-parameter.
But the $T$ parameter constraints require further attention. One
way to handle this \cite{Agashe:2003zs, Agashe:2006at} is by enlarging 
the gauge symmetry 
in the bulk to $SU(3)_c \times  SU(2)_L \times SU(2)_R \times
U(1)_y$ which is an enlarged custodial symmetry which is broken on
the IR brane to recover the SM gauge group. It turns out that the
corrections to the $T$ parameter coming from the dangerous KK gauge
boson sector can be tamed using this custodial symmetry and by
a judicious choice of the fermion representations under the extended
gauge group it is also possible to rein in the non-oblique $Z \rightarrow
b \bar b$ corrections and eventually the bound on the lightest
KK gauge boson mode comes down to about 3 TeV \cite{Davoudiasl:2009cd, 
	Iyer:2015ywa}. 

An alternate proposal to address the issue of the $T$ parameter 
\cite{Cabrer:2010si,Cabrer:2011fb} uses
a deformed metric near the IR brane along with moving the Higgs scalar
into the bulk. 
For this setup, the function $A(y)$ in Eq.~(\ref{e6.11}) is then modified to
\begin{equation}
	A(y)=k y -\frac{1}{\nu^2}\log(1-\frac{y}{y_s})
\end{equation}
The UV brane, similar to the RS setup, is located at $y=0$. The IR brane is however located at $y=y_1$ with the position of the singularity ($y=y_s$) located behind the IR brane at $y_s=y_1+\Delta$, where $\Delta \sim \frac{1}{k}$. $y_1$ is determined by demanding the solution to the hierarchy problem which requires $A(y_1)\sim 35$. The limit $\nu\rightarrow 0$ reverts to the original RS geometry in Eq.~(\ref{e6.11}). 
The deformation of the metric actually causes the Higgs
field to be moved further away from the IR brane but the gauge boson
KK modes are moved by the deformation towards the IR brane. This 
differential action causes the overlap of the Higgs and KK gauge boson
modes to be reduced and that relaxes the bounds coming from the $T$
parameter and the mass of first KK gauge boson mode in this model
can be as small as 1.5 TeV \cite{Iyer:2015ywa}. 

\section{Kaluza-Klein gluons and collider searches}
In typical Bulk RS models, the lightest KK excitations are those of the
gauge bosons and searches for these are likely to be the most
promising probes of such a model. Of these, the KK gluons, because
of their larger couplings, are the most interesting though there
are interesting signals from KK excitations of electroweak
gauge bosons and fermions. 

In the custodial models, the couplings of the $g_{KK}$ to the SM 
states \cite{Agashe:2006hk} are
parametrised in terms of the parameter $\xi \equiv \sqrt{kL} \sim 5$ and
relative to the QCD coupling $g_s$ are given as:
\begin{equation}
	g^{q \bar q g_{KK}} \approx {1 \over \xi} g_s ,
	\hskip15pt
	g^{Q \bar Q^3 g_{KK}} \approx 1  g_s,
	\hskip15pt
	g^{t_R \bar t_R g_{KK}} \approx \xi g_s ,
	\hskip15pt
	g^{g g g_{KK}} = 0 ,
\end{equation}
These denote the coupling of the first Kaluza-Klein mode of the gluon to
light quarks, to the third-generation left-handed doublet, to the
right-handed top quark and to the gluon, respectively. Note that the
$g_{KK}$ couples predominantly to the right-handed top quark
and, consequently, the $g_{KK} \rightarrow t \bar t$ branching ratio is
more than 90\%. For the deformed metric the couplings are similar
except for overall factors to be discussed later.

Also one sees that the coupling of the $g_{KK}$ to
the zero-mode Standard Model gluons vanishes because of the orthonormality
of the Kaluza-Klein modes. This means that, at leading order, $g_{KK}$
production takes place via annihilation of light quarks, to which the
coupling of the $g_{KK}$ is suppressed and, consequently, the cross-section
is small especially since electroweak precision constraints require the mass
of the $g_{KK}$ to be not less than 2-3 TeV. The produced $g_{KK}$ decays 
into a $t \bar t$ pair and
this tiny cross-section has to compete with a huge QCD $t \bar t$ production
background. The fact that this is a resonant cross-section helps somewhat
but then the $g_{KK}$ has a very large width, owing to its strong coupling
to the tops, and so the resonant bump is not sharp but rather smeared out.
The fact that the $g_{KK}$ couples chirally to the tops is, however,
an advantage and a forward-backward asymmetry to pick out the signal
may be used. However, in a $pp$ machine like the LHC this is not easy.
Finally, since the $t$ and the $\bar t$ come from the decay of a heavy
object not less than 2-3 TeV in mass, they are highly boosted. These boosted
top jets can effectively be used as signal discriminant \cite{Agashe:2006hk, 
	Lillie:2007yh}.

Nonetheless, given the smallness of the cross-section it becomes important
to look for other production mechanisms for the $g_{KK}$, especially
the ones which have gluon initial states. The associated production of
$g_{KK}$ with a $t\bar t$ pair leading to distinctive four-top final
states (with two boosted tops) is a process that has been studied with
this in mind \cite{Guchait:2007jd,Lillie:2007hd,Pomarol:2008bh,Servant:2010zza}. The production of $g_{KK}$ 
through top loops has also been considered \cite{Allanach:2009vz} 
though the loop-contributions are greatly suppressed. Model independent 
analysis for the searches of colour octet states (colorons) were considered in \cite{Chivukula:2011ng,Chivukula:2013xla} \footnote{The phenomenology of such models have been studied in \cite{Chivukula:2013xka,Chivukula:2014rka,Chivukula:2015kua}.}.

\section{Associated jet production}
\begin{figure}[!t]
	\begin{center}
		\includegraphics[origin=rb,width=6.2in]{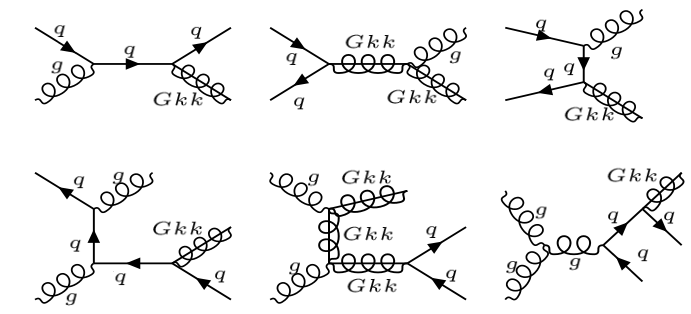}
	\end{center}
	\caption{\it The subprocesses contributing to $g_{KK}$ production in association with partons. Top row is the $g_{KK}$ production in association with a single parton and the bottom row is the production in association with two partons.
	}
	\protect\label{fig1}
\end{figure}

In this paper, we are proposing  the production of a $g_{KK}$ with associated
jets (a light quark or gluon jet) in order to complement the leading order $g_{KK}$ production and increase 
the sensitivity for the $g_{KK}$ at the LHC. At the parton level, the most important contribution is the one
where $g_{KK}$ is produced in association with a single hard parton. Top row in Fig. \ref{fig1} shows few of the contributing diagrams for this process.
As this appears at an order of $\alpha_s$ higher
than the leading-order $g_{KK}$ production, one would think that the
cross-section is smaller. Effects of PDF do partly compensate for the suppression due to $\alpha_s$. In addition,
the associated jet production process has both $q \bar q$ and $qg$ initial
states and a larger number of sub-process contributions. Consequently, the
cross-section for this process is comparable to the leading-order 
$g_{KK}$ production process. Table \ref{resulttable1} and \ref{resulttable2} gives a comparison for the LO cross-section with the cross-section with associated partons for the RS and deformed RS model respectively.

 \begin{table}[here]
 	
 	\begin{center}
 		\begin{tabular}{|c c c c |}
 			\hline
 			\hline                        
 			$g_{KK}$ mass(GeV)	& LO Cross-section(fb)		&  Associated production (AP)(fb)		& AP/LO \\ [0.4ex]
 			\hline
 			\hline 
 			
 			2500	& 169.1	& 107.6	& 0.636\\
 			3000	& 52.53	& 33.28	& 0.634\\
 			3500	& 17.4	& 10.99	& 0.632\\
 			4000	& 5.993	& 3.723	& 0.621\\
 			4500	& 2.096	& 1.277	& 0.609\\

 			\hline
 		\end{tabular}
 	\end{center}
 	\caption{Comparison for cross-sections for mass range in Normal RS }
 	\label{resulttable1}
 \end{table}
 
 \begin{table}[here]
 	
 	\begin{center}
 		\begin{tabular}{|c c c c |}
 			\hline
 			\hline                        
 			$g_{KK}$ mass(GeV)	& LO Cross-section(fb)		& Associated production (AP)(fb)		& AP/LO \\ [0.4ex]
 			\hline
 			\hline 
 			
 			1500 &	1127	& 695.5	& 0.617\\
 			2000 &	257.8	& 162.1	& 0.629\\
 			2500 &	71.42	& 45.51	& 0.637\\
 			3000 &	22.19	& 14.09	& 0.635\\
 			3500 &	7.362	& 4.617	& 0.627\\

 			\hline
 		\end{tabular}
 	\end{center}
 	\caption{Comparison for cross-sections for mass range in Deformed RS }
 	\label{resulttable2}
 \end{table}
We would like to emphasise that, unlike in Refs. \cite{Mathews:2004xp,Mathews:2005bw,Allanach:2009vz}, we are not doing the full NLO analysis here since we are concentrating on hard scattering processes and include only real emission.

The $g_{KK}$ produced via the diagrams shown in Fig. \ref{fig1} will decay 
predominantly to tops. The tops so produced will decay to a $b$ and a $W$ 
which will finally result in a multi-jet final state with the associated
jet being one of these jets. A detailed analysis of the signal and background,
including hadronisation and jet reconstruction is presented in the following
section. We will, however, make a couple of points here before proceeding
to the discussion of the detailed analysis.

In trying to include final states with more than one jet, we are going to
higher orders in QCD perturbation theory. One may wonder then whether we
need to include virtual corrections and worry about soft and collinear
singularities. We do not have to include these corrections because
we are ensuring hard jets in our process. Moreover, some part of the
infrared issues are handled by the showering in {\tt{PYTHIA}} \cite{Sjostrand:2007gs} (albeit in a
model-dependent way) and by the matching of the hard amplitude computation
with the results of {\tt{PYTHIA}} that we have taken into account in a careful
manner. If there is any residual doubt about our calculational procedure,
it is best to bear the analogy with the Drell-Yan process in mind. The
leading-order Drell-Yan process, like leading-order $g_{KK}$ production,
is a $q \bar q$-initiated process. When we consider the high-$p_T$ Drell-Yan
process, in which we consider a single jet recoiling against the
lepton-antilepton pair, then as in our case, the $qg$ process kicks in.
If we were to ask for two hard jets to be produced in association with
the lepton-antilepton pair, then the $gg$ process also comes in. In
computing these hard-scattering amplitudes contributing to high-$p_T$
Drell-Yan, we do not have to worry about the soft or collinear singularities
because we have ensured that the produced jets are hard. Our process
is similar, at the level of the partonic processes, to high-$p_T$ Drell-Yan
(and differs from it only in the kinematics peculiar to the production of
a heavy particle like the KK gluon) and we are, therefore, justified
in neglecting the soft and collinear singularities in our calculation.

Additionally, we calculated the squared matrix-elements for the subprocesses
shown in Fig. 1 explicitly and computed the cross-sections, using a 
parton-level Monte Carlo. This provided us a quick estimate of 
the cross-section and justified our intuition about its magnitude.
It also provided us a rough comparison with the more detailed analysis
that we did using {\tt{MADGRAPH}} \cite{Alwall:2014hca} and {\tt{PYTHIA}} \cite{Sjostrand:2007gs}.

Again, while we postpone the detailed discussion of the choice of kinematic
cuts to the next section, it is useful to note here some features of the kinematics,
arising due to the production of a very massive object at the LHC energies.
 We are interested in producing a $g_{KK}$, at least 1.5 TeV or so in
mass, in association with a jet. Even at the highest energies at the LHC
now available the sub-process centre-of-mass energy will be sufficient
only to produce the $g_{KK}$ with small momentum with the $p_T$-balancing 
associated jet also, therefore, possessing a $p_T$ that is not very large.
When the $g_{KK}$ decays into the $t \bar t$ pair, the $t$ and $\bar t$ are produced almost
back-to-back. But the $t$ and the  $\bar t$ so produced will be highly boosted
and give rise to very collimated hadronic decay products. 
The small magnitude of the $g_{KK}$ $p_T$ also implies that the recoiling parton carries small net $p_T$ which 
also poses a challenge in trying to distinguish it from the other hadronic environment. The resultant final state
has a $g_{KK}+$ several jets which could arise from partonic sub-processes with more than one associated parton.
We therefore also take into account the cross-section due to processes with additional partons.
The process with two additional partons arises also from $gg$ initial states (in addition to the $q\bar q$ and $qg$ channels) which contribute at the 
lower orders. 
The bottom row in Fig. \ref{fig1} shows the production of $g_{KK}$ in association with two partons.
The cross-section is again not very suppressed. Processes with larger number of partons in association will follow the usual perturbation theory pattern and are expected to be suppressed.
Further processes with more than two partons will have a tendency to produce softer jets which may not pass the cuts.
 We have checked the cross-section for $g_{KK}$ with one, two and three additional partons in {\tt{MADGRAPH}}
and find the above expectation to be borne out. We therefore include the contribution of one parton and two parton sub-processes to the signal cross-section.

\section{Signal and Background simulations}
The signal is characterised by a massive RS KK-gluon ($g_{KK}$) produced in association with hard partons. The $g_{KK}$ then decays dominantly into a $\bar t t$  pair leading to the following signal topology
\begin{equation}
	p~p\rightarrow g_{KK}+a ~\quad\text{or}\quad ~p~p\rightarrow g_{KK}+a+b
	\label{NLO}
\end{equation}
where $a,b$ are partons and $g_{KK}\rightarrow t\bar t$.
The requirement of an additional hard parton does not change the signal. 
This process is at the next to leading order in comparison to following leading order (LO) topology
\begin{equation}
	p~p\rightarrow g_{KK}\rightarrow \bar t t
	\label{LO}
\end{equation}

In comparison the $g_{KK}$ produced in a LO process will be almost at rest.  The parton-level amplitudes for both the signal and the background were generated using {\tt{MADGRAPH}} \cite{Alwall:2014hca} at 13 TeV centre of mass energy using parton distribution function {\tt{NNLO1}} \cite{Ball:2012cx}. The model files have been generated using {\tt{FEYNRULES}} \cite{Alloul:2013bka}. The signal events were generated by adding the following amplitudes:
\begin{equation}
	\mathcal{M}_{signal}=\mathcal{M}(p p\rightarrow g_{KK})+\mathcal{M}(p p\rightarrow g_{KK}+a )+\mathcal{M}(p p\rightarrow g_{KK}+a+b )
\label{signal}
\end{equation}

The most dominant backgrounds are $t\bar t +$ \textit{jets} and QCD.
The events for the former is generated by adding the following amplitudes at the parton level:
\begin{eqnarray}
	\mathcal{M}_{bg}=\mathcal{M}(p p\rightarrow t \bar t )+\mathcal{M}(p p\rightarrow t \bar t+a )+\mathcal{M}(p p\rightarrow t \bar t+a+b)
\end{eqnarray}

The QCD background is simulated by adding the following amplitudes at the parton level
\begin{eqnarray}
\mathcal{M}_{bg}=\mathcal{M}(p p\rightarrow a+b )+\mathcal{M}(p p\rightarrow a+b+c)
\end{eqnarray}

The associated partons for the $t\bar t$ and the signal are required to have a minimum transverse momentum of 50 GeV. A softer cut enhances the background cross-section and hence not desirable. The signal cross-section on the other hand is not drastically affected as the associated parton is expected to have a fairly large transverse momentum from the recoiling $g_{KK}$.

We need the `hard jet' from the matrix elements in the second and third terms in the right hand side  of Eq.~(\ref{signal}). The `soft jets' are modelled by emissions off $p p\rightarrow g_{KK}$ in the first term generated by {\tt{PYTHIA 8}} \cite{Sjostrand:2007gs}.
This was done by matching the {\tt{MADGRAPH}} output to {\tt{PYTHIA 8}} using {\tt{MLM}} \cite{Mangano:2002ea} matching scheme. 
Subsequently, after showering and hadronisation using {\tt{PYTHIA 8}} we apply the following selection criteria to extract the signal:\\

\textit{Jet Selection:} Assuming hadronic decay of the top, the final state is characterised by at least 6 partons (including 2 $b$ partons). Jets are reconstructed from these partons by employing  {\tt{FASTJET}} \cite{Cacciari:2006sm,Cacciari:2011ma} using the $anti-k_T$ \cite{Cacciari:2008gp} clustering algorithm and setting the jet radius parameter $R=0.4$. We accept only those jets which satisfy $|\eta|<2.5$ and $p_T>50$ GeV.\\

\begin{figure}[!t]
	\begin{center}
		\begin{tabular}{c}
			\includegraphics[width=8.2cm]{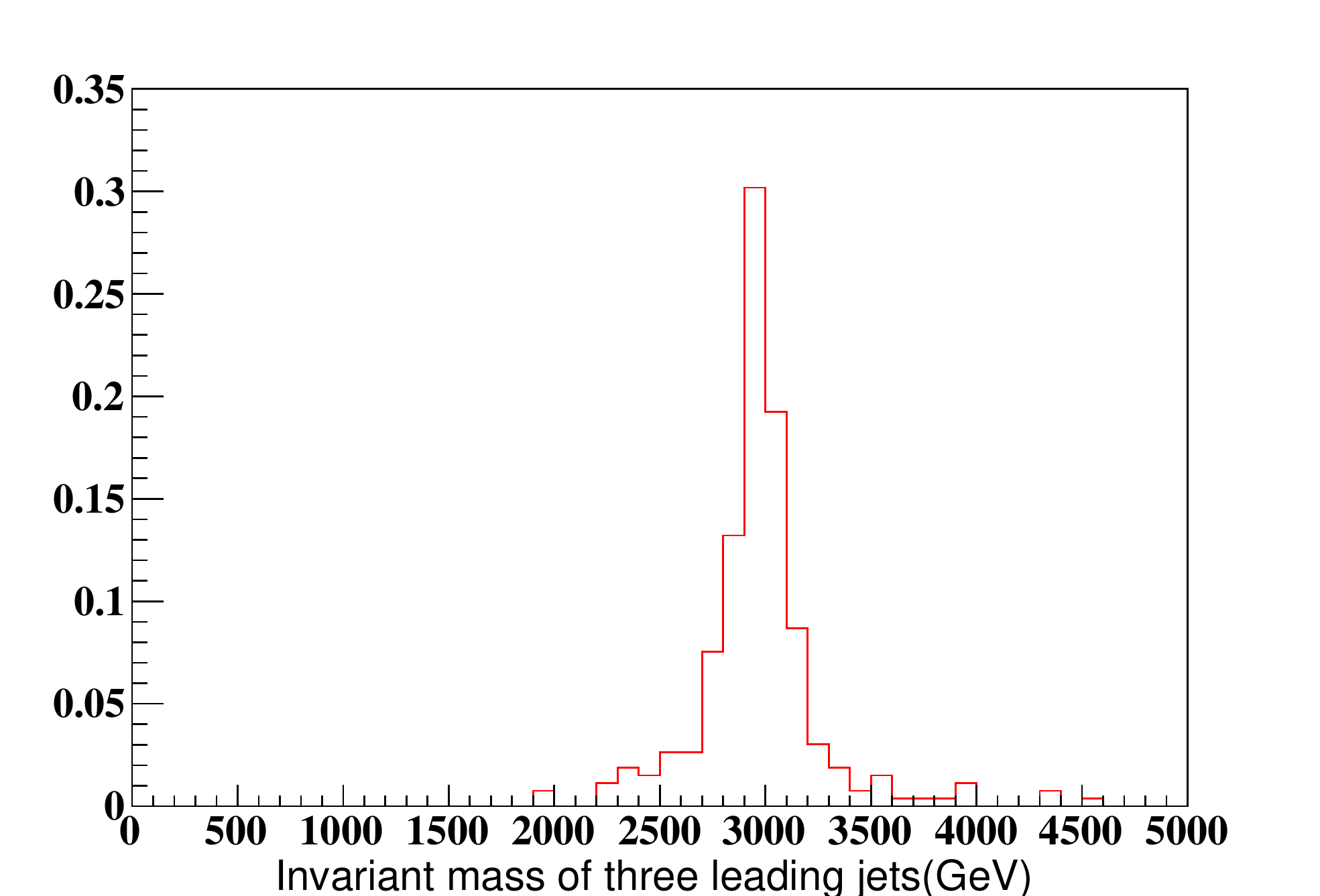}\\
		\end{tabular}
	\end{center}
	\caption{Reconstructed mass of the $g_{KK}$ from the three leading jets where the mass of $g_{KK}$ is chosen to be 3 TeV.}
	\label{KKGrecon}
\end{figure}

\textit{Event Selection:} 
Leptons are required to satisfy $p_T>20$ GeV and $\eta<2.5$.
Since we are considering a hadronic decay channel for both the tops, only events with no leptons are accepted. 
In order to reconstruct the $g_{KK}$ mass from the top decay products, each event is further required to have a minimum of 3  jets. An additional condition on a maximum of 5 jets is imposed which is helpful to reduce qcd background
Due to the heavy mass of the $g_{KK}$, the decay products of $t$ are likely to be reconstructed in a single jet.
Alternatively, one could have imposed the requirement of a minimum $4$ normal jets \footnote{Jets which are not identified as $b-$ jets.} and 2 $b-$ jets.
However the requirement of a minimum of two $b$ jets in each event results in depletion of signal events due to $b$ tagging efficiency.
As a result, we do not attempt to segregate the $b-$ like jets to those coming from the $W-$ decay and $g_{KK}$ mass is reconstructed from the 3 leading jets in the event. Fig.~\ref{KKGrecon} shows the invariant mass of 3 leading jets in each event demonstrating a distinct peaking behaviour at 3 TeV which is the mass chosen for the $g_{KK}$. \\

\begin{figure}[!t]
	\begin{center}
		\begin{tabular}{c}
			\includegraphics[width=9.2cm]{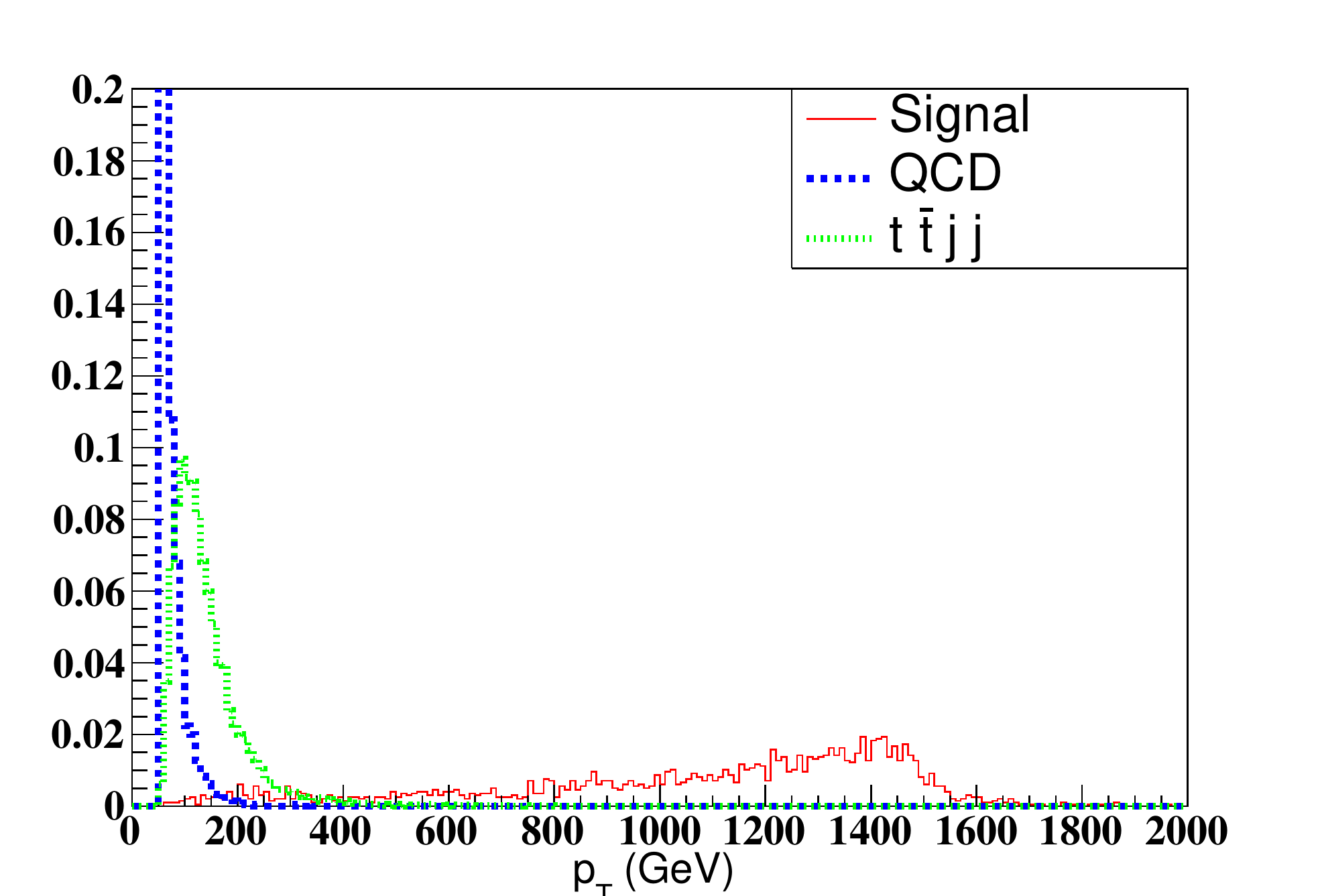}
		\end{tabular}
	\end{center}
	\caption{$p_T$ distribution for the leading jet for the signal (red) and background (blue) for $m_{g_{KK}}=3$ TeV.}
	\label{ptdist}
\end{figure}

\textit{Background rejection:}  
For the signal, the top pairs and hence its decay products are likely to be highly boosted owing to large $g_{KK}$ mass. As a result the leading jet ($j_0$) is likely to have a very large $p_T$ in comparison to  that of the background.
Fig. \ref{ptdist} shows the $p_T$ distribution for the leading jet for the background (blue and green) and the signal (red).
We give a hard cut of 1100 GeV on the leading jet. To increase the statistics for the background in this high $p_T$ region, background events are simulated near the tail end of the $p_T$ spectrum.  This is implemented by demanding the sum of $p_T$ of the associated partons for background simulation is 1250 GeV.
Further, each event is required to have a minimum of two  a maximum of 8 jets. 
Inspite of such a hard cut on the leading jet, contamination due to QCD processes is non-zero due to its significant production cross-section. Traditional kinematic cuts prove insufficient to get a significant signal sensitivity. 

\begin{figure}[!t]
	\begin{center}
		\begin{tabular}{cc}	
			\includegraphics[width=8.2cm]{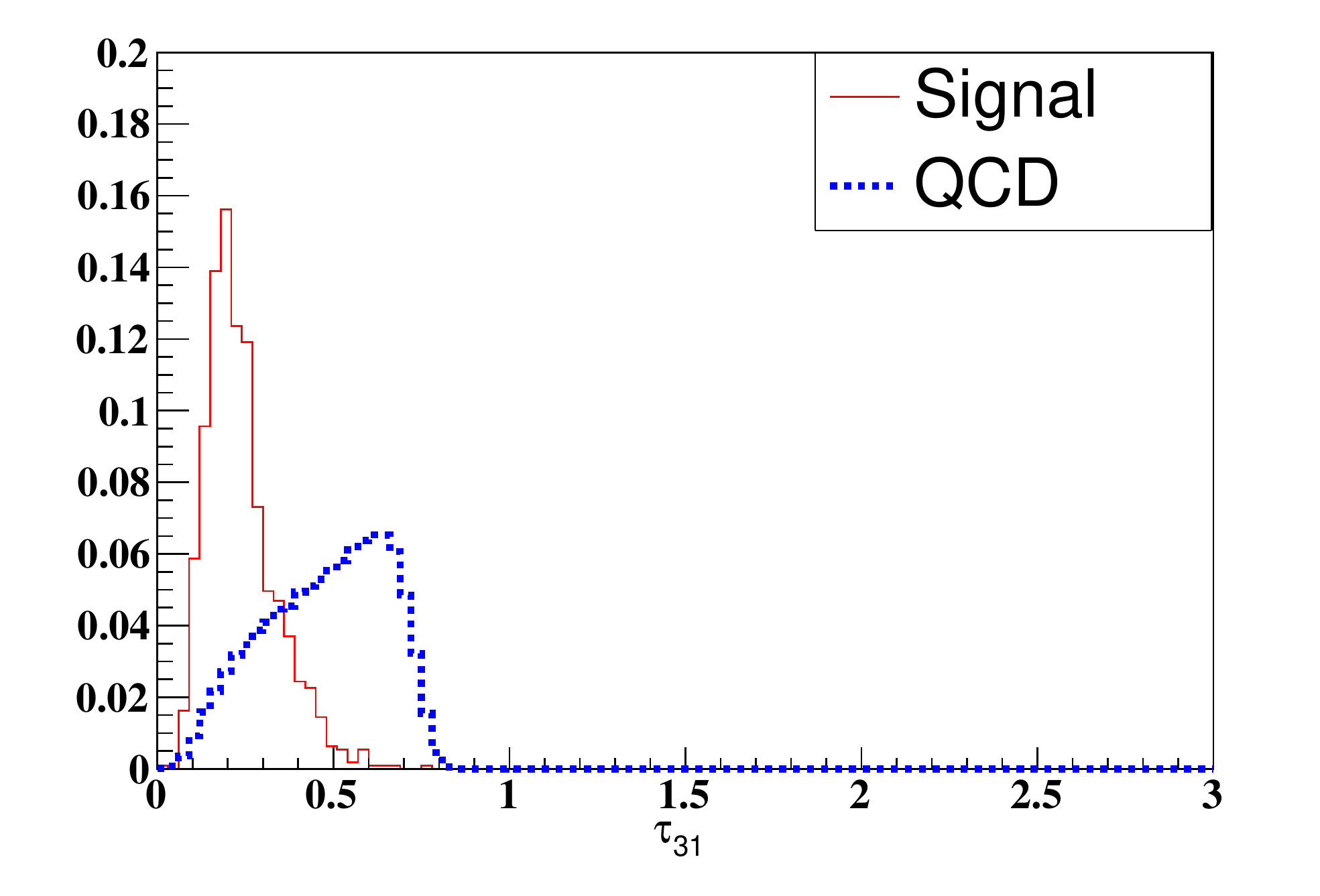}&	\includegraphics[width=8.2cm]{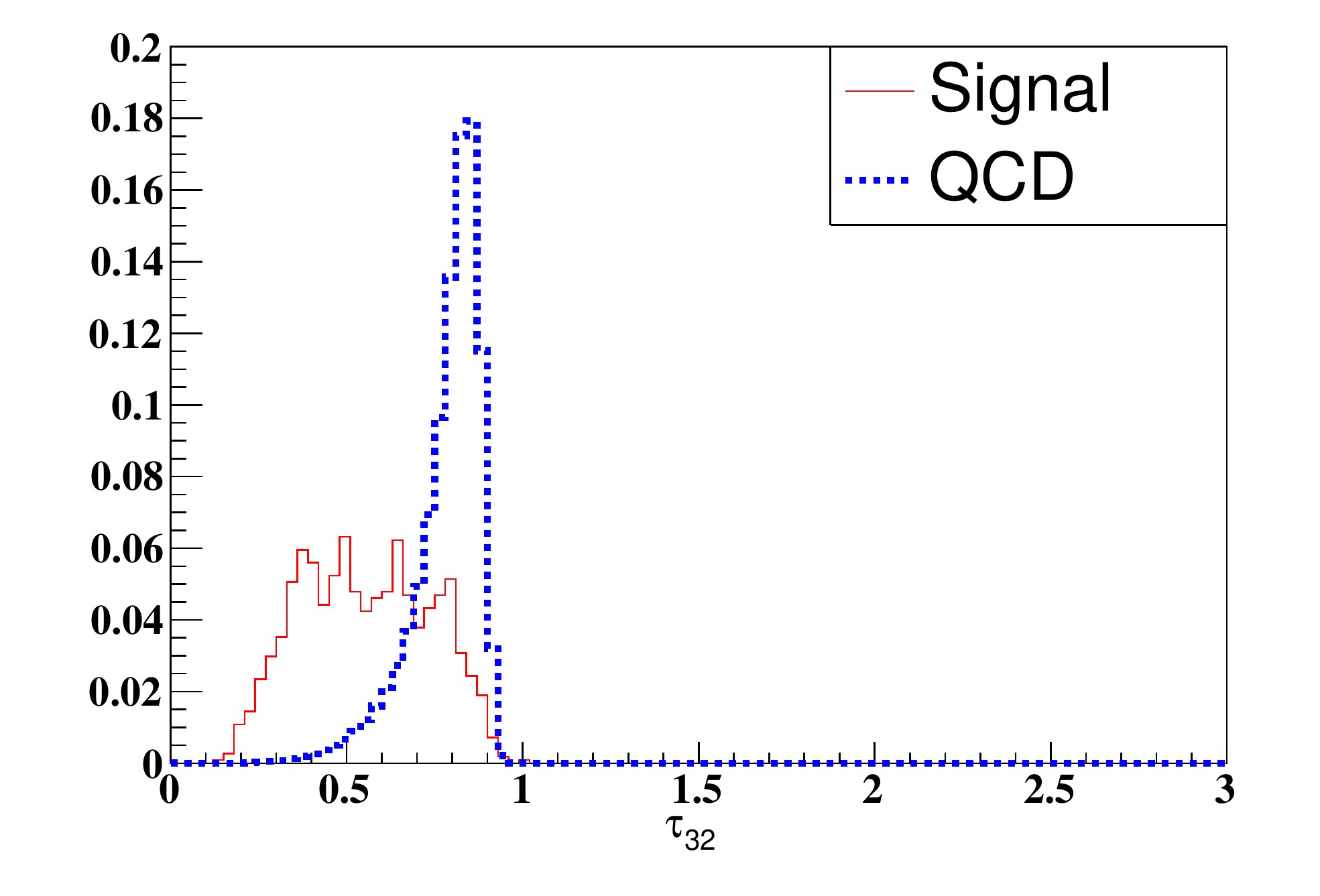}\\
		\end{tabular}
	\end{center}
	\caption{Subjettiness ratio $\tau_{31}$ on the left and $\tau_{32}$ on the right for both signal and background. }
	\label{subjettiness}
\end{figure}

In this case, studying the jet substructure is extremely useful in eliminating the background to a great extent. It was pointed out earlier that the leading jet for the signal is primarily composed of top decay products. As a result it has a three-lobed structure. This is in sharp contrast to QCD which is one-lobed.  
A useful way to quantify the number of lobes for a given jet is by considering a variable called \textit{N-subjettiness} \cite{Thaler:2010tr,Thaler:2011gf} defined as:
\begin{equation}
\tau_N=\frac{\sum_k p_{T_k}\times \text{min}\left( \Delta R_{1k},\Delta R_{2k}\ldots \Delta R_{Nk}\right)}{\sum p_{T_k} \times R }
\label{nsubjettiness}
\end{equation}
where $k$ runs over the jet constituents and $p_{T_k}$ is the transverse momentum for the $k-th$ constituent. 
In the above definition, assuming there are $N$ candidate subjets,
$\Delta R_{lk}$ is the distance in the rapidity-azimuth plane between the $k-th$ constituent and the $l-th$ candidate subjet.
For a jet with $N-$ distinct lobes of energy, all the radiation inside the jet will be aligned along their direction, which is the same as the direction of the candidate subjet. As a result each constituent of the jet
will be clustered with one of the subjets and one can expect the $\text{min}\left( \Delta R_{1k},\Delta R_{2k}\ldots \Delta R_{Nk}\right)$ to be closer to zero.
Thus, $\tau_N\rightarrow 0$ for $N-$ lobe configuration while $\tau_{N-1}>\tau_N$ for $N>1$. In this case $\tau_{N+1}$ is expected to be comparable to $\tau_N$. 
For QCD $\tau_1$ is small while for the signal $\tau_3$ is small.
A useful application is to consider ratios $\tau_{N+1,N}=\tau_{N+1}/\tau_N$. For the scenario under consideration, we evaluate $\tau_{32}$ and $\tau_{31}$. Both these ratios are expected to be smaller for the signal than QCD as shown in Fig. \ref{subjettiness}.
One of the most dominant background to the signal is $t\bar t+$ jets production via SM processes due to its significant production cross-section. 
It is therefore essential to suppress this background significantly to have sufficient signal over background efficiency for luminosities attainable in the near future. We select $\tau_{31}<0.35$ and $\tau_{32}<0.35$ for the leading jet. Further we also impose a cut of $\tau_{21}<0.6$ on the sub-leading jet.
 
\textbf{Results:}

Table \ref{resulttable} gives the summary of the number of events passing various cuts at each level for both the signal and background. 
These results correspond to $m_{g_{KK}}=3$ TeV. The cuts are optimised for this particular mass.
As explained earlier events with zero leptons are accepted to facilitate hadronic decay of the top. A hard cut on the transverse momentum of the leading jet ($j_0$) drastically reduces the $t\bar t$ background without affecting the signal significantly.
With this set of cuts, $\sim 5\sigma$ sensitivity can be obtained for a minimum luminosity of 90 fb$^{-1}$ for $m_{g_{KK}}$ = 3 TeV.

\begin{table}[!t]
	\begin{center}
		\begin{tabular}{|c c c c c|}
			\hline
			\hline                        
			Sr.No & Cuts & QCD($p^{sum}_T>1250$) &$t\bar t j j$($p^{sum}_T>1250$)& Signal events\\ [0.5ex]
			\hline
			\hline 
			1 & Given number of Events & $10^7$ &$10^5$ &10000\\
			2&Cross-section(fb)&$720\times 10^3$ &132&94.2 \\
			3 & $n_{lepton}=0$  & 9203793 & 39386&3122\\
			4 &  $p^{j_0}_T>$1100 GeV & 233520  & 14051&1363\\
			5&Subjettiness cut& 262&218 &265\\
			6 & $|m_{g_{KK}}-3000| < 80$ GeV&48   & 55&112\\

			\hline
		\end{tabular}
	\end{center}
	\caption{Cut flow table for $m_{g_{KK}}=3$ TeV.}
	\label{resulttable}
\end{table}

\begin{figure}[here]
	\begin{center}
		\begin{tabular}{cc}
			\includegraphics[width=9.2cm]{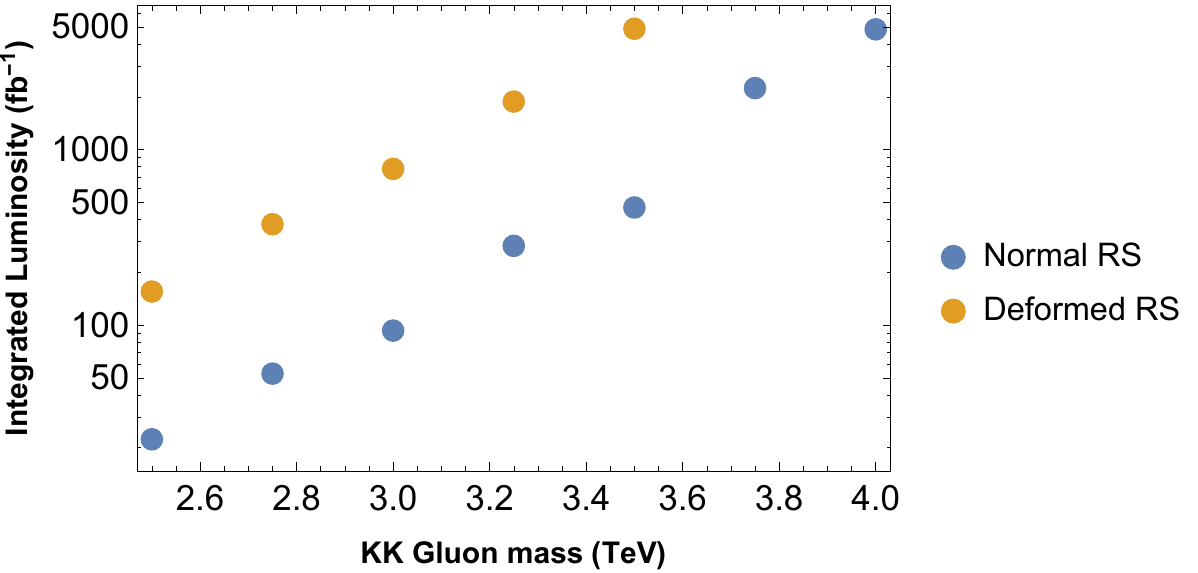}&\includegraphics[width=7.2cm]{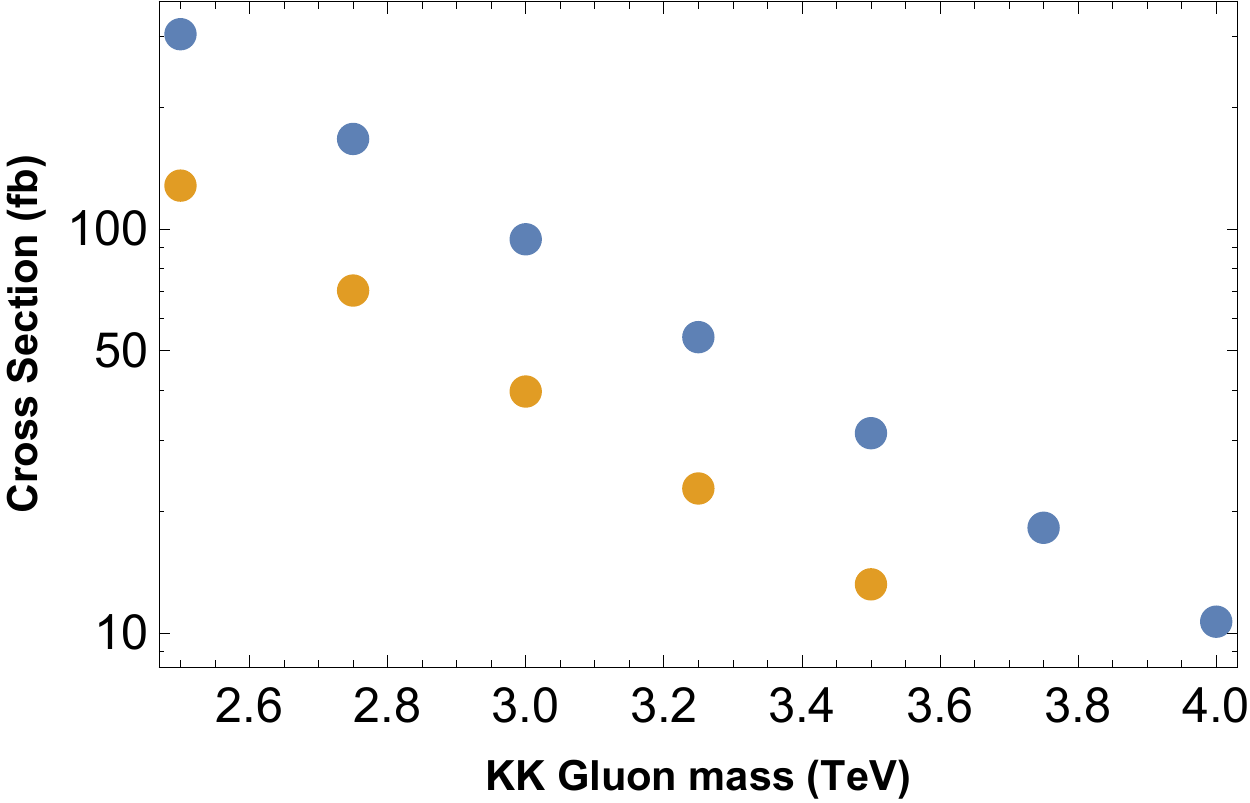}\\
		\end{tabular}
	\end{center}
	\caption{Minimum luminosity required for a 5$\sigma$ sensitivity for normal RS (blue) and deformed RS (orange). The right plot shows the production cross-section for the different masses.}
	\label{reach}
\end{figure}

We repeat the analysis above for different masses of the $g_{KK}$ and we follow exactly the same pattern of cuts as in Table \ref{resulttable}. The background events are simulated differently for different masses of $g_{KK}$ since the fifth line in Table \ref{resulttable} uses the cut around the mass of the KK-gluon and presents out results for both normal and deformed RS model. For the normal RS model we assume 92\% branching fraction into $t\bar t$ pair, while for the deformed model we assume 83\% branching fraction \cite{Carmona:2011ib}. 
Left plot in Fig. \ref{reach} presents the minimum luminosity required for a $5\sigma$ signal sensitivity for the different masses for both normal RS and deformed RS models.
Owing to constraints from precision electroweak data we do not consider masses below 2.5 TeV for normal RS model. Due to their larger production cross-section, the lower masses (indicated by blue points in the figure) have better sensitivity in terms of early discovery prospects.

Lower masses can be admitted in a deformed RS model. 
For the deformed model we choose $\nu=0.4$ while the curvature radius is chosen to be $L_1=0.2/k$ \cite{Carmona:2011ib,deBlas:2012qf}.
This scenario however suffers from reduced production cross-section owing to the smaller coupling of the $g_{KK}$ to the lighter quarks which is roughly $0.13g_s$.

For both scenarios, we use a very hard cut on the transverse momentum of the leading jet, $p_T^{j_0}>1100$ GeV in Table \ref{resulttable}. Since the $p_T$ of the leading jet is $\sim m_{g_{KK}}/2$, this cut is more effective for the heavier masses as compared to the lighter masses. While this depletes majority of the signal points for 2 TeV KK gluon, this is helpful in depleting the background to a great extent. 

\section{Conclusion:}
Search for KK excitations of the gluons in warped framework is interesting due to their relatively large production cross-section in comparison to other KK states. Their masses are, however, strongly constrained due to limits from electroweak precision data. This necessitates the need for an effective strategy to probe relatively heavy states in the Run II of LHC. We consider a process where the $g_{KK}$ is produced in association with jets.  We consider a simple set of cuts to extract the signal from the background. The signal is characterised by highly collimated leading jets owing to the massive nature of $g_{KK}$. Cuts as strong as 900 GeV is imposed on the $p_T$ of the leading jet  without adversely affecting the number of signal sensitivity. We present results for both the normal RS model and the deformed RS model. In normal RS, $g_{KK}$ as heavy as 3 TeV can be probed in the Run II of the LHC with luminosities $\sim 90$ fb$^{-1}$ while masses as heavy as 4 TeV can be accessed in the HL-LHC. For the deformed case masses $\sim 2.5$ TeV are accessible in the current run of LHC. The sensitivity to 3 TeV masses can be probed in the HL-LHC. 

While this simple cut based analysis is highly effective, it would be interesting to find alternate strategies so that one can explore the heavy mass regime more efficiently. The observation of highly collimated decay products of either top coming from the $g_{KK}$ provides strong motivation to study the jet-substructure in greater detail and constitutes work for the future.

\section{Acknowledgements}
We would like to thank Monoranjan Guchait for the extensive discussions and immensely useful inputs on the collider analysis. We are also grateful to him for the going through the manuscript in detail and for his suggestions on improving the quality of the draft. AI would like to thank Amit Chakraborty for discussions on simulations. NM would like to thank the Department of Theoretical Physics, TIFR for computational resources. We would also like to acknowledge the contributions of Sophie Renner to the
initial stages of this work.

\textit{The authors would like to dedicate this paper to the memory of Guido Altarelli.}


\bibliographystyle{ieeetr}

\bibliography{mybib1.bib}

\end{document}